\documentclass[a4paper,3p,times]{elsarticle}
\usepackage{graphics,graphicx,dcolumn,bm,fleqn,epic,eepic,float}
\usepackage{amssymb,amsmath,multirow,rotate,color,float,times}

\usepackage{color}
\usepackage{soul}                    
\definecolor{red}{rgb}{1,0,0}
\definecolor{green}{rgb}{0,1,0}
\definecolor{blue}{rgb}{0,0,1}

\newcommand{\bite}{\begin{itemize}}
\newcommand{\eite}{\end{itemize}}
\newcommand{\benu}{\begin{enumerate}}
\newcommand{\eenu}{\end{enumerate}}
\newcommand{\bverb}{\begin{verbatim}}
\newcommand{\everb}{\end{verbatim}}

\newcommand{\beq}{\begin{equation}}
\newcommand{\eeq}{\end{equation}}
\newcommand{\barr}{\begin{array}}
\newcommand{\earr}{\end{array}}

\begin{document}

\title{Self-organized criticality in a network of\\
       economic agents with finite consumption}
\author[closer,cftc]{Jo\~ao P.~da Cruz}
\ead{joao.cruz@closer.pt}
\author[cftc,fcl]{Pedro G.~Lind} 
\ead{plind@cii.fc.ul.pt}

\address[closer]{Closer Consultoria Lda, Avenida Engenheiro Duarte Pacheco,
             Torre 2, 14$^o$-C, 1070-102 Lisboa, Portugal}
\address[cftc]{Center for Theoretical and Computational Physics, 
             University of Lisbon,
             Av.~Prof.~Gama Pinto 2, 1649-003 Lisbon, Portugal}
\address[fcl]{Departamento de F\'{\i}sica, Faculdade de Ci\^encias 
             da Universidade de Lisboa, 1649-003 Lisboa, Portugal} 

\begin{abstract}
We introduce a minimal agent
model to explain the emergence of heavy\--tailed 
return distributions as a result of self-organized criticality.
The model assumes that agents trade their economic outputs with each other 
composing a complex network of agents and connections.
Further, the incoming degree of an agent is proportional to the 
demand on its goods, while its outgoing degree is proportional to the supply.
The model considers a collection of economic agents which are attracted 
to establish connections among them to make exchange at a price form by 
supply and demand. 
With our model we are able to reproduce the evolution of the return
of macroscopic quantities (indices) and to correctly retrieve the non-trivial
exponent value characterizing the amplitude of drops in several
indices in financial markets, 
relating it to the underlying topology of connections.
The distribution of drops in empirical data is obtained by counting
the number of successive time-steps for which a decrease in the index
value is observed. All eight financial indexes show an exponent $m\sim 5/2$.
Finally, we present mean-field calculations of the critical exponents, 
and of the scaling relation $m=\tfrac{3}{2}\gamma-1$
between the exponent $m$ for the distribution of drops
and the topological exponent $\gamma$ for the degree distribution.
\end{abstract}



\maketitle

\section{Introduction: Observing SOC in financial data}


The application of statistical physics to finance and economy
was boosted in the last decades, particularly with the
analysis of financial data in 1973 by Black, Scholes and 
Merton\cite{blackscholesmerton}, explaining the price evolution
in an organized market. More recently\cite{friedrich2000,ghashghaie96}, 
such application found important developments
with the introduction of procedures for quantitatively describing 
financial data, by explicitly deriving a Fokker-Planck equation 
for the empirical probability distributions which catches the typical 
non-Gaussian heavy tails of financial time-series, across scales.
However, as Mandelbrot pointed out\cite{mandelbrot},
other typical features are observed in the variation of prices, namely
scale invariance behavior\cite{mantegna1995}, 
which cannot be explained by means of a
cascade model. The modeling of financial index dynamics 
and other complex phenomena taking into account non-Gaussianity and scale 
invariant behavior has been indicated as an appealing problem to address 
in the scope of finance analysis and statistical 
physics\cite{mantegna1995,kiyono}, particularly in what concerns
the emergence of Self-organized Criticality (SOC)\cite{bak87} in 
financial markets.

SOC is typically observed in slowly-driven non-equilibrium systems with 
extended degrees of freedom and a high level of nonlinearity. 
Many individual examples have been identified since the Bak, Tang and
Wiesenfeld (BTW) model in the pioneer paper in Ref.~\cite{bak87}, describing
the occurrence the emergence of power-laws in different systems.
This simple model shows avalanches of arbitrary sizes in one cellular 
automaton mimicking a sand pile, where at each sand cell is toppled to
other cells when a certain threshold level is reached. The toppling may 
trigger subsequent topplings composing what one can interpret as an 
avalanche.

Surprisingly, such cascade of topplings was reported to be observed in
other contexts, assuming the proper interpretation, and even when not
completely accepted, SOC become established as a strong candidate for
explaining the phenomenology of such systems.
For example, 
evolution of species seems to lay around a self-organized critical state 
of periods with almost no mutations and then periods with arbitrarily 
large sequences of mutations exhibiting a power-law 
distribution\cite{bak93}.
Another important example is earthquake statistics, which was already
known to obey Gutenberg-Richter law\cite{gutenberg}, a power-law which agrees
with SOC phenomenology.
The fluctuations in economic systems, described for instances by financial
markets indices or prices were also reported by Mandelbrot\cite{mandelbrot},
Mantegna and Stanley\cite{mantegna1995} and other to show  
scale invariance behavior.
But to the best of our knowledge there is still no clear connection
between the power-laws seen in financial markets and SOC.

Using an agent model to characterize two groups of traders, Lux and
Marchesi presented some evidence that scaling in finance emerges as
a result of the interactions between individual market 
agents\cite{lux1999}. 
Agent-based models of financial markets have been intensively studied.
For a review see Ref.~\cite{LuxReview2007} and references therein.
It has been claimed that agent models are able to complement the 
approach borrowed from social sciences, where from one theory a
specific model is derived and applied to empirical data\cite{mossPNAS}.
An agent model is typically constructed by starting to identify the
statistical features in the data of interest and then to implement
the necessary ingredients in our model in order to generate data
with the same features.

In this paper, 
we aim to show that two fundamental assumptions 
in Economics are sufficient for the emergence of a self-organized 
critical state in the social environments where humans trade, an 
environment that we call henceforth the financial system and that 
we map into a set of interconnected agents.
Further, we will show that the under such assumptions, the model
is also able to reproduce the scaling features observed in empirical 
data.

The first assumption is that humans
are attracted to each other to exchange labor due to biological specialization 
that made the species more efficient when each specimen could perform the tasks
to which is more capable\cite{lipseybook}. The second is that the amount of 
labor exchanged by each of the two involved agents in an economic relation 
is ruled 
by the law of supply and demand\cite{lipseybook}. Based on these principles, 
which are straightforwardly translated to the physical context,
we will show that the assumption for the system to be in equilibrium cannot be 
sustained, differently from the Brownian particle approach.
Since labor must be produced in order that agents can establish labor 
exchanges, on each agent there is an ``energy'' dissipation that 
must finite. 
Inserted in a system 
where agents are impelled to create more economic exchanges then, physically, 
this configuration corresponds to a self-organized critical system\cite{bak87}.
In other words, the self-organized critical state emerges due to finite 
``energy'' consumption in economic agents.
Though the term energy used in the economical context is not 
the same as physical energy, since it does not necessarily satisfy the
thermodynamic constraints, we will use the term in the economical context
only. For example, human labor is assumed as ``energy'' delivered by
the agent to a neighbor, which rewards the agent with an energy that the
agent accumulates. The balance between the labor produced for the neighbors
and the reward received from them may be positive (agent profits) or 
negative (agent accumulates debt). For simplicity, we omit henceforth the 
quotation marks.

Since the main concern in risk management deals with the 
distribution of the so-called drops in financial market indices, we 
restrict ourselves to that side of the distribution.
In a financial market index, one drop is defined as the decrease
of the index from one time-step to the next one.
Each drop on financial markets results from the economical crash
of one or more agents, being represented by the collapse of those agents.
Such local crashes appear at no particular moment and may cascade 
into an avalanche of arbitrary size, i.e.~a succession of an
arbitrary large number of local economical crashes, taken
as a global economical crisis.

We start in section \ref{modelo} 
by introducing the minimal model that translates the two 
fundamental assumptions described above into a two body problem without 
additional assumptions.   
In the section \ref{finnet}, we frame the two body problem in the many body 
problem assuming that agents connect each other by preferential 
attachment\cite{barabasi} as expected from other human networks to show 
the critical behavior of the system with a typical exponent around
$5/2$,  characterizing the critical behavior in
the system, and that the return distribution 
probability depends directly on the topology of the network.
We show that our model reproduces the typical behavior of financial indices 
drops, in particular showing that empirical indices have similar exponent 
values, around $5/2$.
The exponent $5/2$ is derived quantitatively based in branching processes
and mean-field assumptions and show that it depends linearly on the 
underlying complex network exponent.
In section \ref{conclusions} we draw the conclusions.
    
\section{Minimal Model for Economic Relations}
\label{modelo}

\begin{figure}[b]
\begin{center}
\includegraphics*[width=6.0cm,angle=0]{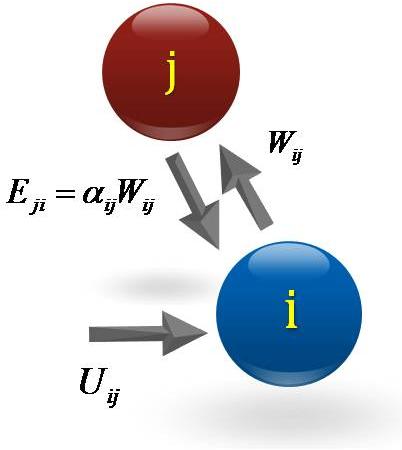}%
\includegraphics*[width=6.0cm,angle=0]{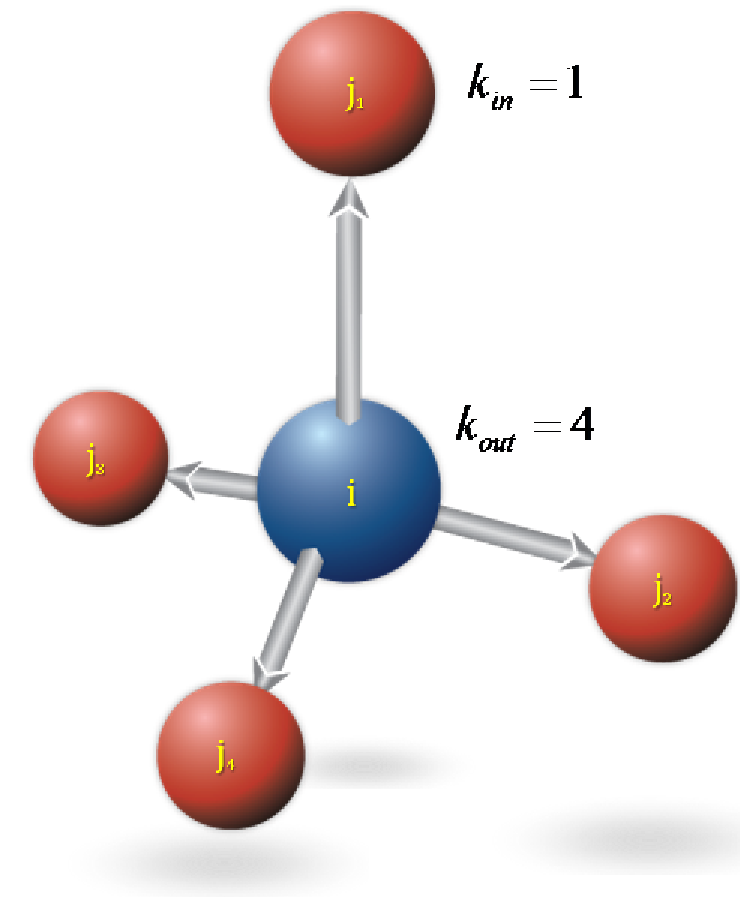}
\end{center}
\caption{\protect 
  {\bf Left:}
  Illustration of an economical connection between two
  agents $i$ and $j$. Agent $i$ transfer labor $W_{ij}$ to agent $j$ receiving
  an energy $E_{ij}=\alpha_{ij}W_{ij}$ where $\alpha_{ij}$ measures how well the
  labor is rewarded: agent $i$ has an outgoing 
  connection (production) and agent $j$ an incoming connection (consumption).
  This interaction attributes to agent $i$ an amount
  of ``internal energy'' $U_{ij}=W_{ij}-E_{ij}$ that can be summed up over all
  agents connections
  up to a threshold $U_{th}$ beyond which it is distributed
  among the neighbors (see text).
  {\bf Right:} The transfer of labor is done according to a preferential
  attachment scheme: the agent prefers to work for agents which have
  already a large number of labor connections to them.}
\label{fig1}
\end{figure}

In this section we introduce an agent-model that
considers a collection of $N$ agents operating as energy 
transducers into the economic space in the form of labor. 
Agents, labeled $i$ and $j$ in
Fig.~\ref{fig1}
(left), establish among them bi-directed connections characterized
as follows.
When agent $i$ delivers labor $W_{ij}$ to agent $j$ it receives in return a 
proportional amount of agent $j$ labor, $E_{ij}=\alpha_{ij}W_{ij}$.
Both quantities $E_{ij}$ and $W_{ij}$ can be regarded as forms of energy,
in the economic space. 
Henceforth we consider all interchange of energy in units of $W_{ij}=1$.

The factor $\alpha_{ij}$ is a (dimensionless) measure of the labor price,
defined as
\begin{equation}
\alpha_{ij} = \frac{2}{1+e^{-(k_{out,i}-k_{in,j})}}
\label{alpha}
\end{equation}
where $k_{out,i}$ and $k_{in,j}$ are the number of
outgoing connections of agent $i$
and the number incoming connections of agent $j$, respectively 
(see Fig.~\ref{fig1}). 
A large (small) $k_{in,j}$ indicates a large (small) supply
for agent $i$ and a large (small) $k_{out,i}$ indicates a large (small) 
demand of agent $i$. 

When $\alpha_{ij}=1$, agent $j$ returns to agent $i$ the same amount of labor 
it receives from agent $i$.
This happens when $k_{in,j} = k_{out,i}$, yielding $W_{ij}=E_{ij}$,
i.e.~when there is local balance between supply and demand.
This value $\alpha_{ij}=1$ is the middle value 
between the asymptotic limits  
$k_{in,j}\gg k_{out,i}$ ($\alpha_{ij}\sim 0$) 
and $k_{in,j}\ll k_{out,i}$ ($\alpha_{ij}\sim 2$) 
which satisfy basic economic principles\cite{lipseybook}.
Namely, in the limit $k_{in,j}\gg k_{out,i}$, the labor of
agent $j$ is in much greater demand than the labor of $i$,
and thus agent $j$ is in a position to pay $i$ very
little in return.
In other words, for $\alpha_{ij}<1$ 
the labor of agent $i$ is paid by $j$ below 
the amount of energy $W_{ij}$ it delivers, i.e.~agent $i$ loses from 
the connection (trade) and loses a certain amount of energy, 
$U_{ij}=W_{ij}-E_{ij}>0$.
For $\alpha_{ij}>1$ the opposite occurs, i.e.~
when $k_{in,j} < k_{out,i}$, agent $i$ has more supply of labor
than agent $j$ has demands for its labor.
In the limit $k_{in,j} \ll k_{out,i}$, the value of $\alpha_{ij}$ 
could in principle be any finite value larger than one.
To guarantee an equal range length for both the situation 
when agent $i$ profits from agent $j$ and the situation
when agent $i$ looses from agent $j$, we consider the
range $\alpha_{ij}\in [0,2]$, which implies the limit $\alpha_{ij}=2$
for $k_{in,j} \ll k_{out,i}$ as can be seen from Eq.~(\ref{alpha}).

The energy balance for each trade, $W_{ij}-E_{ij}$, translates into
an energy balance for each node that takes into account
all outgoing and incoming connections the agent has:
$U_i=\sum_{j\in {\cal V}_{out,i}}(W_{ij}-E_{ij})+\sum_{j\in {\cal V}_{in,i}}
(E_{ji}-W_{ji})$ where ${\cal V}_{out,i}$ and ${\cal V}_{in,i}$
are the outgoing and incoming vicinity of agent $i$, with 
$k_{out,i}$ and $k_{in,i}$ neighbors respectively.

We also assume that each agent $i$ chooses its neighbors 
according to a preferential
attachment scheme, following the Barab\'asi-Albert-Jeong
method\cite{barabasi}. 
This scheme is as follows:
one starts with a small amount of agents totally interconnected, 
and adds iteratively one agent with one connection to one of the 
previous agents, chosen from a probability 
function proportional to their number of connections.
Thus, agents having a large amount of connections are more likely
to be chosen for a new connection than other agents.

Here, this topology underlies the 
empirical observation in economic-like systems that 
agents are more likely to deliver their work to agents 
receiving already significantly amount of work.
See Ref.~\cite{barabasi} for several examples such as
the Internet and the airport network among other.
See Fig.~\ref{fig1}.

Since we drop the assumption of an equilibrium system, these 
particles we call agents can experiment the boundaries of the 
system and leave it if they go too far from the expected general 
equilibrium conditions\cite{vonnewmann}. Thus, as agents build new 
economic connections 
the factors $\alpha_{ij}$ and the energy of agents change.
The number of incoming connections 
(consumption) can overcome the outgoing ones (production)
up to a certain threshold  $U_{th,i}$ that is related to how much the 
system allows the agent to accumulate debt.
Following standard economical reasoning\cite{refMerton},
the amount of debt an agent may accumulate is directly related 
with the volume of its overall business, i.e., with the broadness of 
its influence in the system: 
If one agent produces more than another, it is reasonable to
expect that the systems allows him to accumulate more debt.
Representing 
this influence by the turnover $T_i=k_{out,i}+k_{in,i}$, we fix a threshold
$d_{th}=U_{th,i}/T_i$ against which we compare the measure $d_i=U_i/T_i$.

Under these assumptions we consider that when $d_i < d_{th}$ the agent
collapses and an avalanche takes place. This collapse induces the
removal of all the $k_{in,i}$ consumption connections of agent $i$ from
the system, implying  
\begin{subequations}
\begin{eqnarray}
U_{i} & \rightarrow & U _ {i}+\sum_{j\in {\cal V}_{in,i}}(1-\alpha_{ji}) \label{eq:avalanche1} \\
T_{i}  & \rightarrow & T_{i} - k_{in,i} \label{eq:avalanche2} \\
U_{j}  & \rightarrow & U _ {j} - (1-\alpha_{ji}) \label{eq:avalanche3} \\
T_{j}  & \rightarrow & T_{j} - 1 ,\label{eq:avalanche4}
\end{eqnarray} 
\label{eq:avalanche}
\end{subequations}
where $j$ labels each neighbor of agent $i$.
The collapse of agent $i$ generates a new energy balance on agent $j$.
If $j$ does not collapse, the avalanche stops and is saved as an 
avalanche of size $s=1$. If $j$ collapses the avalanche continues to
spread to the next neighbors.

\section{Emergence of SOC in Financial Networks and Empirical Financial 
indices}
\label{finnet}

Having presented the model, we show next that
under the above assumptions
the system remains at a critical state.
To this end, we make use of a mean-field approach for
such a system.
The factors $\alpha_{ij}$ 
are substituted by the average value 
$\alpha=\langle \alpha_{ij}\rangle$ yielding for each agent
\begin{equation}
U_i = (1-\alpha)(k_{out,i}-k_{in,i})  .
\label{interenergy}
\end{equation}

Due to the preferential attachment scheme\cite{barabasi}, 
in the initial state of the system, the outgoing
connections follow a $\delta$-distribution $P_{out}(k)=\delta(k-k_{out})$
and the incoming connections follow a scale-free distribution
$P_{in}(k)=k_{in}^{-\gamma_{in}}$,
where $\gamma_{in}$ is the exponent of the degree distribution.
As the system evolves, the number of agents remains constant
but at each event-time $n$ one new connection joining two agents
is introduced, with both agents independently chosen according 
to the preferential attachment scheme mentioned above.
Thus, through evolution both consumption and production networks
are pushed to a degree distribution of the form $P(k)\sim K_0k^{-\gamma}$, 
where $K_0$ is the initial number of outgoing connections a node has.

As equated in Eqs.(\ref{eq:avalanche}), a collapsing agent $i$ 
has all its consumption connections removed from its neighbors.
Thus, the collapse of a neighbor $j$ occurs if 
$k_{out,j}-k_{in,j} > d_{th}(k_{out,j}+k_{in,j})$ and 
$k_{out,j}-1-k_{in,j} \leq d_{th}(k_{out,j}-1+ k_{in,j})$, yielding 
\begin{equation}
\omega k_{in,j} < k_{out,j} \leq \omega k_{in,j}+1 
\label{eq:eqavalanche_3}
\end{equation} 
with $\omega=\frac{1+d_{th}}{1-d_{th}}$.
Taking a collapsing node, the probability $P_{br}$ for a neighbor to also 
collapse is the probability for the above condition to be fulfilled.
Since all connections are formed by preferential attachment, 
\begin{equation}
P_{br}(k_{in,j}) = P(\omega k_{in,j} < k_{out,j} \leq \omega k_{in,j}+1) \approx K_0(\omega k_{in,j})^{-\gamma}  .
\label{eq:pbr}
\end{equation}
\begin{figure}[htb]
\begin{center}
\includegraphics*[width=15.0cm,angle=0]{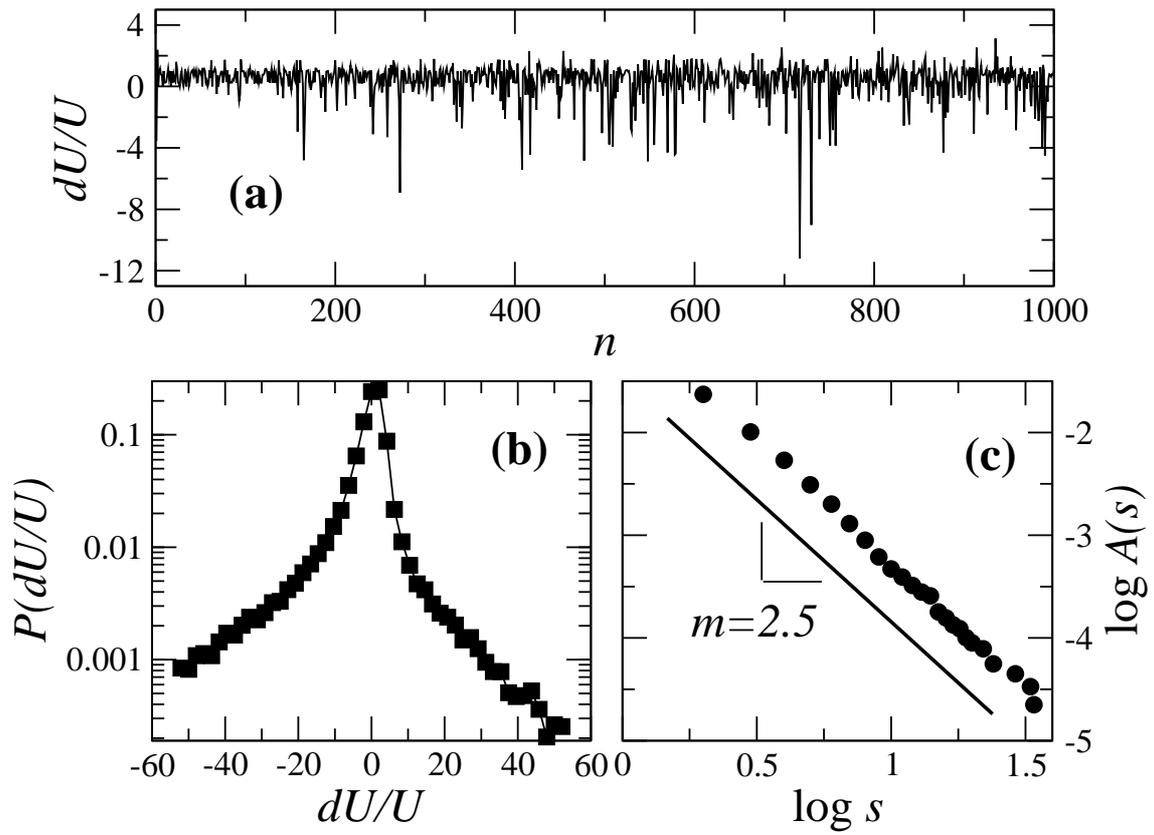}
\end{center}
\caption{\protect 
  {\bf (a)} Evolution of the variation of the total internal energy, 
  which shows {\bf (b)} probability density function and {\bf (c)}
  avalanche size distribution similar to the ones observed for
  empirical data (compare with Fig.~\ref{fig1}).
  Here $N=1000$, $K_0=1$ and $W_{ij}=1$ for all $i$ and $j$.}
\label{fig2}
\end{figure}
\begin{figure}[bth]
\begin{center}
\includegraphics*[width=15.0cm,angle=0]{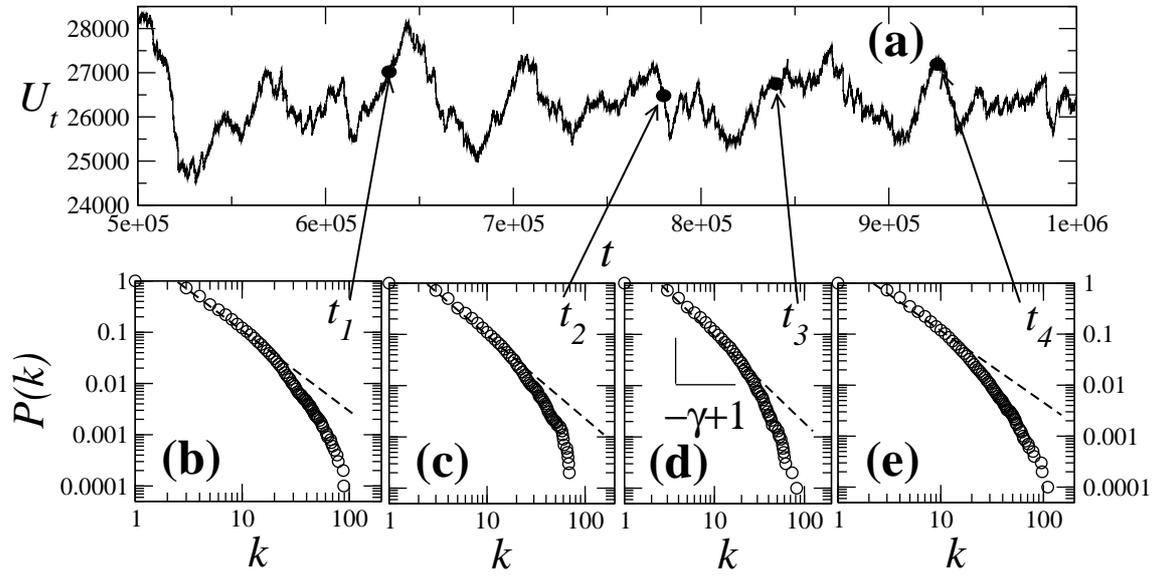}
\end{center}
\caption{\protect 
         {\bf (a)} Time series of the total internal energy $U$ of the 
         agents. For the four instants $t_1$, $t_2$, $t_3$ and $t_4$ we
         show the observed cumulative degree distribution 
         $P(k^{\ast}\ge k)\propto k^{-\gamma+1}$ yielding exponents
         {\bf (b)} $\gamma=2.6\pm 0.5$,
         {\bf (c)} $\gamma=2.7\pm 0.5$,
         {\bf (d)} $\gamma=2.6\pm 0.5$,
         {\bf (e)} $\gamma=2.6\pm 0.5$,
         all of them according to the theoretical prediction 
         (check Eq.~(\ref{eq:avalanchefinalacum})).
         The deviations from the power-law for large $k$ are due
         to the avalanches (crisis) in the system (see text).}
\label{fig3}
\end{figure}
\begin{figure}[htb]
\begin{center}
\includegraphics*[width=15.0cm,angle=0]{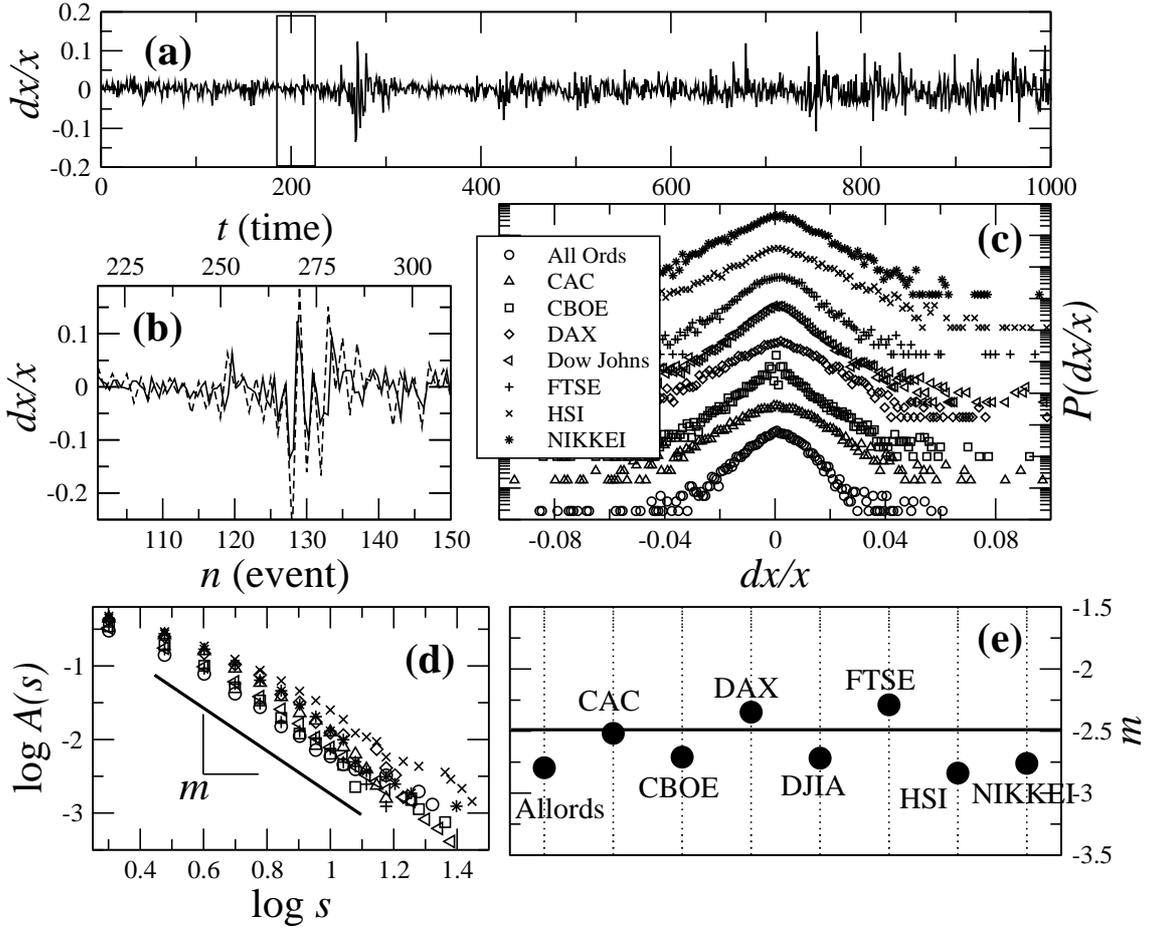}
\end{center}
\caption{\protect 
    {\bf (a)} Time evolution of the logarithmic returns of the DJIA index 
(partial). The box is zoomed in {\bf (b)} to emphasize the contrast 
between real data (solid line) throughout time $t$ and the succession of
events (dashed line) where the variation changes sign.
{\bf (c)} Probability density functions for some important financial 
indices, including interest rate options (CBOE).
{\bf (d)} Distribution of avalanche sizes detected throughout the evolution 
of each financial index showing critical behavior with 
{\bf (e)} an exponent $m$ approximately invariant and similar to the
one obtained for branching processes and to our model.}
\label{fig4}
\end{figure}

To know if one collapsing agent triggers an avalanche one needs
to estimate the expected number of neighbors that the 
agent brings to collapse, due to its own collapse.
If this expected number would be smaller than one, the systems would need
to consume an infinite amount of energy from the environment.
If the expected number would be larger than one, the entire system
would typically be extinct by one large avalanche.
The avalanche of collapsing agents form a branching process and 
assuming that the system cannot consume an infinite amount 
of energy from the environment and that the system survives the avalanches, 
the expected value of collapsing agents from a starting one
must be equal to one\cite{harris} and therefore
\begin{equation}
\sum _{k_{in,i}=1} ^{\infty}{k_{in,i} P (k_{in,i})P _{br}(k_{in,i})}=1
\label{critcond}
\end{equation}
which yields 
$\sum _{k_{in,i}=1} ^{\infty}{ k_{in,i}^{-\gamma}}=\left ( 
\frac{\omega}{K_0}\right )^\gamma$,
i.e.~the system remains in the critical state\cite{harris}  as
\begin{equation}
\omega^{\gamma}=K_0^{\gamma}\zeta(2\gamma-1)
\label{eq:valoresperadoki}
\end{equation}
where $\zeta$ is the Riemann zeta-function.  
Condition (\ref{eq:valoresperadoki}) closes our model, 
relating economic 
growth ($K_0$), topology ($\gamma$) and the allowed level of debt($\omega$).
From Otter's theorem\cite{otter49} for branching processes 
the distribution for the avalanche size expressed as number of agents $r$ is
given by $P(r)\propto r ^{-\frac{3}{2}}$. Since the energy of our system is
expressed as connection number, the number of collapsed agents
in an avalanche 
is given by $r=NK_0(\omega k_{in}))^{-\gamma}$, where $N$
is the total number 
of agents in the system. Therefore, the degree distribution 
is given by $P(k)\propto k^{-\frac{3}{2}\gamma}$ and the avalanche 
size distribution reads
\begin{equation}
P(k\geq s)\propto \int _{s} ^{+\infty} {k^{-\frac{3}{2}\gamma} dk} \propto s^{-\frac{3}{2}\gamma+1} \equiv s^{-m} .
\label{eq:avalanchefinalacum}
\end{equation}
Equation (\ref{eq:avalanchefinalacum}) relates the exponent characterizing
the network topology with the exponent taken from the avalanches, using
first principles in economy theory. 

Next we show that for the typical values of $\gamma$ 
found in empirical networks,
our model reproduces the values of $m$ predicted in 
Eq.~(\ref{eq:avalanchefinalacum}).
For that, we define a 
macroscopic quantity for the internal energy, which
accounts for all outgoing connections in the system at each time step:
\begin{equation}
  U=\sum _{i=1}^{N}{\sum_{j\in {\cal V}_{out,i}}} {(W_{ij}-E_{ij})} .
 \label{eq:totalenergy}
\end{equation}
The quantity $U$ varies through time, and its evolution reflects 
the development or fail of the underlying economy, similar to
a finance index. 
Alternatively, $U$ can be calculated 
from the incoming connections.
This macroscopic quantity will be used
to characterize the state of our economic-like system.

Figure \ref{fig2}a shows a sketch of the evolution of a typical
time-series for the logarithmic returns 
$dU/U$. As can be seen from Fig.~\ref{fig2}b the distribution of
the logarithmic returns is non-Gaussian with the heavy tails observed
in empirical data\cite{kiyono}.
In Fig.~\ref{fig2}c the cumulative distribution $A(s)$
of the avalanche size $s$ is plotted showing a power-law whose fit yields 
$A(s)\sim s^{-m}$ with an exponent $m=2.51$ ($R^2=0.99$).
Looking again to Eq.~(\ref{eq:avalanchefinalacum}) and 
borrowing from the literature\cite{barabasi}
the values of $\gamma$ of empirical networks
which lay typically in the interval $[2.1,2.7]$,
one concludes that the exponent should take typical
values $m\in [2.15,3.05]$ which agrees with the results from our model.

Figure \ref{fig3} shows how the overall index dynamic 
emerges from the underlying network mechanics. 
As agents connect each other by preferential attachment, the topology 
of the system is pushed to a power law degree distribution. 
On the other hand, avalanches push the system away from it. 
Thus, the system undergoes a structural fluctuation that generates a 
fat tail distribution of the index, expressed by 
Eq.~(\ref{eq:avalanchefinalacum}) and shown in Fig.~\ref{fig4}, 
rather than a Gaussian one.
Figure \ref{fig3}a shows a typical set of successive $U_t$ values 
taken from our model. The cumulative degree distribution $P(k)$ at the
marked instants $t_1$-$t_4$ are shown in Fig.~\ref{fig3}b-\ref{fig3}e.
The dashed lines guide the eye for the scaling behavior observed
at the lower part of the degree spectrum.
In all cases $\gamma\sim 2.6$. Varying the threshold
one observes other values for exponent $\gamma$ (not shown).
For large degree $k$ the distribution deviates from
the power-law, due to the drops of connections for 
agents experiencing an economical crash. 
Nonetheless, 
the slope $-\gamma+1$ of the dashed 
line yields values in the predicted range,
within numerical errors. 
 
Next we address the observation that 
the results obtained from our model in Fig.~\ref{fig2} do agree with
the analysis done on eight main financial indices, as shown in 
Fig.~\ref{fig4}. 


The time-series of the logarithmic returns (Fig.~\ref{fig4}a) must 
first be mapped in a series of events.
One event is defined as a (typically small) set of successive instants 
in the original time-series having the same derivative sign, either 
positive (monotonically increasing values) or negative
(monotonically decreasing values). 
Each time the derivative changes sign a new event starts. 
Figure \ref{fig4}b shows an zoom of the original time
series in Fig.~\ref{fig4}a (solid line) with the corresponding
series of events. In the continuous limit, events would correspond
to the instants in the time-series with vanishing first-derivative.
Further,
to be comparable to empirical series, we consider in our
analysis a sampling of data which takes one measure of the original
series from the system each five iterations.

The non-Gaussian distributions of the logarithmic returns 
(Fig.~\ref{fig4}c) were extracted from the logarithm
returns of the original series of each index, as in
Ref.~\cite{kiyono}. The characteristic heavy 
tails observed in \cite{kiyono} are observed for short time
lag (hours or smaller), where in Fig.~\ref{fig4}c the daily 
closure values are considered.
The power-law behavior of the avalanche size 
(Fig.~\ref{fig4}d) is indeed similar to the simulated results.
Moreover the exponents $m$ have all approximate values, plotted in
Fig.~\ref{fig4}e, around the simulated value $m=2.51$ (solid line),
and predicted by Eq.~(\ref{eq:avalanchefinalacum}).

All empirical indices are sampled daily but in different time periods.
For FTSE $6498$ days in London stock market were considered,
starting on April 2nd 1984 and ending on December 18th 2009.
For DJIA $20395$ days in New York stock market were considered,
starting on October 1st and ending on December 18th 2009.
For DAX $4815$ days in Frankfurt stock market were considered,
starting on November 26th 1990 and ending on December 18th 2009.
For CAC $5003$ days in Paris stock market were considered,
starting on March 5th 1990 and ending on December 18th 2009.
For ALLORDS $6555$ days in Australian stock market were considered,
starting on August 3rd 1984 and ending on June 30th 2010.
For HSI $5701$ days were considered in Hong Kong stock market, 
starting on December 31st 1986 and ending on December 18th 2009.
For NIKKEI $6386$ days were analyzed in Tokyo stock market, 
starting on January 4th 1984 and ending on December 18th 2009.
For CBOE IR10Y $12116$ days in Chicago derivative market,
starting on January 2nd 1962 and ending on June 30th 2010.

\section{Conclusions}
\label{conclusions}

In this paper, we have showed that, based only on first principles of
economic theory and assuming that agents form an open system of economic 
connections organized by preferential attachment mechanisms, 
one is able to reach the distribution of drops observed in financial 
markets indices, including stocks and interest rate options. 
Assuming that the preferential attachment mechanism is part of the growing 
of economic connections, the resulting self-similar topology allows us to 
assume that the total economy system may present a similar
topological structure as the sub-economy around financial markets and, 
thus, market indices can be taken as good proxies for 
the total economy.

We presented evidence that the distribution of drops in financial indices 
reflects the degree distribution taken from the trading network of the 
economical agents.
In other words, the topology and structure of economic-like networks
strongly influences the frequency and amplitude of economical crisis.
Quantitatively, we showed that 
the two exponents characterizing the degree distribution and the
distribution of drops, respectively, obey a scaling relation.
Further, we showed how the scaling relation can be derived from a mean-field 
approach,
assuming that the avalanche is a branching process of the economical
agents and measuring its amplitude from the expected number of trading
connections that are lost.

It should be noticed that 
the mean-field approach does not provide insight on local variations
of both exponents.
Differences between the different indices are also related to the 
different social-economic realities beneath them, including e.g.~growing 
periods or crisis.
For example, despite the fact that, in general, all economies have the 
same behavior, during a growing period the structure of the economic 
network shows a broader degree distribution (see e.g.~Fig.~\ref{fig3}e)
corresponding to a higher value of the proxy  (financial index).

The results averaged for sufficiently long time show
that the numerical model here introduced
reproduces the exponent $m\sim 5/2$ for the distribution of the drops 
observed in empirical financial indices.
The value of the exponent $m$ gives indications of how large 
are crisis in the corresponding economy. Subsequent work that we
are now finishing and will publish elsewhere shows that the exponent
m is bounded by $[2,3.5]$.

Two final remarks are due here. 
First, as stated in the introduction,
we only dealt 
with drops, occurring for a lower threshold of the difference between 
consumption and production at one single agent. Though, the interpretation 
of the total ``energy'' in the system as a financial index for the market 
can only be closed if the so-called ``booms'' are also considered in the
evolution of the financial proxy.
The booms were not incorporated in our model, 
since we were concerned with risk management.
To incorporate them
an additional threshold in the consumption would be needed.

Second,
our results show that the full topology underlying economic-like
systems plays an important role in the evolution of the proxies 
characterizing the economical state. In the particular context of financial
networks, one may rise the question of how successful are 
directives and policies, such as Basel III, when they aim to 
avoid crisis through directives such as increasing minimum capital level
in each financial agent.
By translating this minimum capital level as a local threshold for each
agent of a financial network, our model can be adapted to simulate future
scenarios of the overall financial stability subjected to different
thresholds.
We have show that surprisingly the raise of such capital
levels could drive the entire financial system into a state 
where larger crisis are more probable.
This application of the model here introduced will be described in detail
elsewhere\cite{epl}, with a thorough discussion of the results.

\section*{Acknowledgments}
\label{akn}

The authors thank M.~Haase, F.~Raischel and N.R.~Bernardino for
useful discussions. 
The authors thank financial support from PEst-OE/FIS/UI0618/2011, 
PGL thanks  {\it Funda\c{c}\~ao para a Ci\^encia e a Tecnologia -- 
Ci\^encia 2007} for financial support.


\end{document}